\documentclass[aps, prl, reprint, superscriptaddress, nofootinbib]{revtex4-2}

\usepackage[english]{babel}
\usepackage{import}
\usepackage{enumerate}
\usepackage{graphicx}
\usepackage{color}
\usepackage[dvipsnames]{xcolor}
\usepackage{appendix}
\usepackage[applemac]{inputenc} 
\usepackage{fontenc}  
\usepackage{amsmath}
\usepackage{comment} 
\usepackage{subfigure}
\usepackage{dsfont} 
\usepackage{float}
\usepackage[normalem]{ulem}
\definecolor{linkcolor}{rgb}{0,0,0.6}		
\definecolor{bleu}{HTML}{1732a6}
\usepackage[colorlinks=true, pdfstartview=FitV, linkcolor= linkcolor, citecolor= linkcolor, urlcolor= linkcolor, hyperindex=true, hyperfigures=false]{hyperref}
\newcommand{\ket}[1]{| #1 \rangle}
\newcommand{\bra}[1]{\langle #1 |}
\newcommand{\beq}{\begin{equation}}
\newcommand{\eeq}{\end{equation}}
\newcommand{\bea}{\begin{eqnarray}}
\newcommand{\eea}{\end{eqnarray}}
\newcommand{\ba}{\begin{array}}
\newcommand{\ea}{\end{array}}

\newcommand{\bef}{\begin{figure}}
\newcommand{\eef}{\end{figure}}

\begin{document}

\author{Maria Maffei}
\affiliation{Universit\'e Grenoble Alpes, CNRS, Grenoble INP, Institut N\'eel, 38000 Grenoble, France }

\author{Patrice A. Camati}
\affiliation{Universit\'e Grenoble Alpes, CNRS, Grenoble INP, Institut N\'eel, 38000 Grenoble, France }

\author{Alexia Auff\`eves}
\affiliation{Universit\'e Grenoble Alpes, CNRS, Grenoble INP, Institut N\'eel, 38000 Grenoble, France }

\title{Probing non-classical light fields with energetic witnesses \\ in Waveguide Quantum Electro-Dynamics}

\begin{abstract}
We propose an operational scenario to characterize the nature of energy exchanges between two coupled but otherwise isolated quantum systems. Defining work as the component stemming from effective unitary interactions performed by each system on one another, the remnant is stored in the correlations and generally prevents full energy extraction by local operations. Focusing on the case of a qubit coupled to the light field of a waveguide, we establish a bound relating work exchange and local energy extraction when the light statistics is coherent, and that gets violated in the presence of a quantum light pulse. Our results provide operational, energy-based witnesses to probe non-classical resources.

\end{abstract}
\maketitle

With the rise of quantum technologies, understanding the laws governing the exchange of energy and entropy between quantum systems is of crucial importance~\cite{Binder2018}. Work is a powerful concept in the classical world, that features an entropy-preserving, well-controlled energy change~\cite{Callen1998}. In the quantum realm, such energy changes typically take place along unitary interactions, i.e. when the system of interest gets classically driven -- as it is the case for quantum gates. Extending this concept to coupled quantum systems is still exploratory and proves to be a very active field of research nowadays~\cite{Campaioli2018,Weimer2008,Hossein-Nejad2015,Alipour2016,Sparaciari2017,Schroder2018}. As a distinctive feature, coupled quantum systems may get entangled during their interaction. The resulting entanglement entropy challenges the concept of work captured by classical intuitions. 

In \cite{Weimer2008, Hossein-Nejad2015}, a general framework was proposed to analyze the nature of energy flows between two coupled but otherwise isolated quantum  systems. Within this simple partition of the physical world, energy flows split into two components, respectively stemming from the effective unitary driving that each system performs on one another and from the creation of correlations between the two systems. The former (resp. the latter) quantity is identified as the work flow (resp. the correlation energy flow), in agreement with the usual definition of work when one of the systems becomes classical and no entanglement appears.

\begin{figure}[t]
	\begin{center}
\includegraphics[width=0.95\columnwidth]{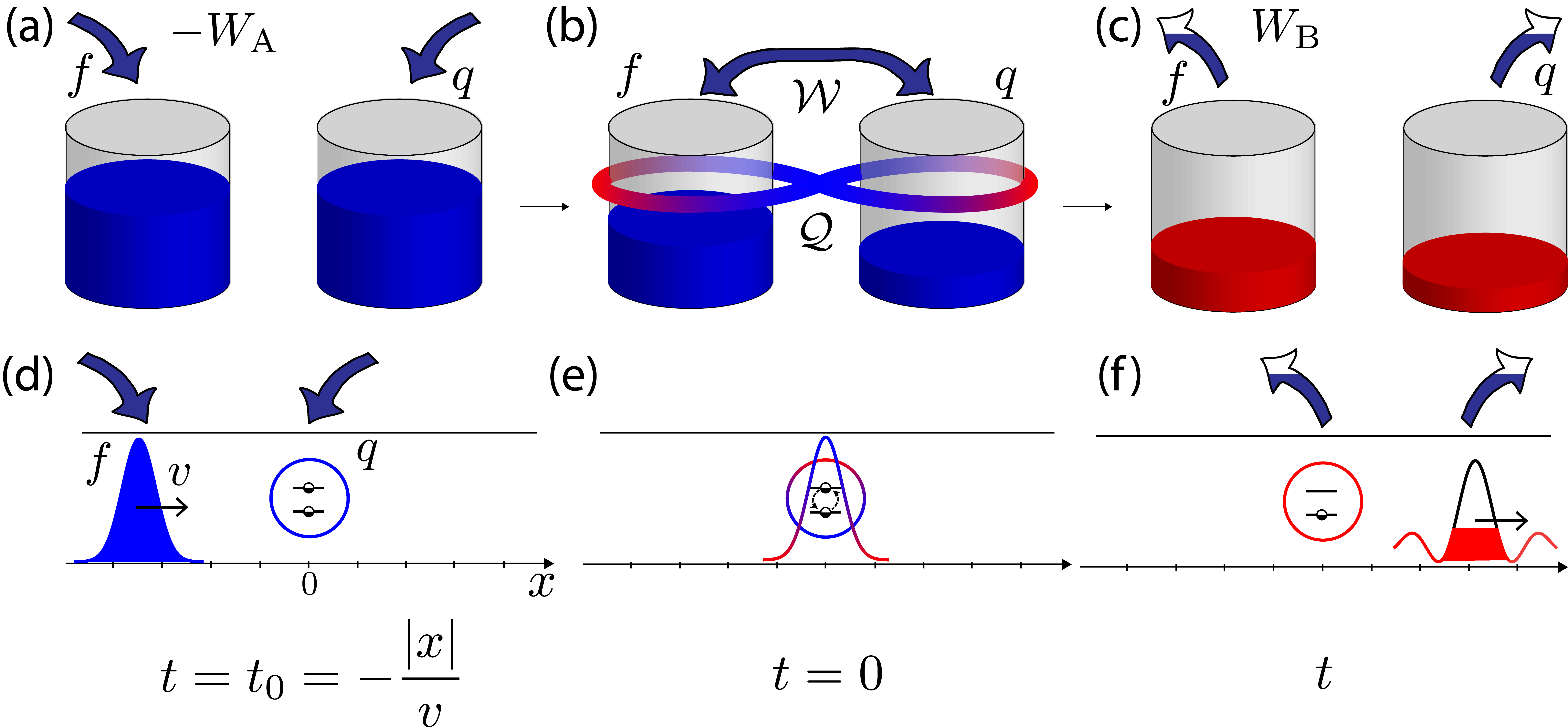}
\caption{(a)-(c) Operational scenario. (a) Alice prepares the initial state of the bipartite system $qf$ at time $t=t_0$, spending the amount of work $-W_\text{A}$ (blue). (b) at time $t=0$ the two subsystems are interacting, exchanging work (blue) and correlation energy (blue and red). (c) at time $t$ Bob locally extracts work from both subsytems, $W_\text{B}$. Due to the correlations, some locally unextractable energy may remain (red). (d)-(f) Waveguide setup. A qubit is located at the position $x=0$ of a 1DWG. (d) At $t=t_0=-|x|/v$, Alice prepares a pulse of light and the qubit in a pure state. The field propagates without dispersion in the region $x<0$ (resp. $x>0$) at the velocity $v$. (e) At $t=0$ the pulse impinges the qubit. (f) At time $t$ the qubit and the pulse are still entangled, preventing full energy extraction by local unitary operations. 
\label{fig1}
}
\end{center}
\end{figure}

In this article, we first propose a scenario where these definitions acquire a transparent operational meaning, see Figs.~\ref{fig1}(a)-(c). Energy is locally injected into each system to prepare a pure product state, then gets delocalized while entanglement builds up. Finally, the invested energy is tentatively recovered, only using local unitary operations. While entanglement generally hinders local energy extraction, full recovery can be reached at any time provided that the two systems solely exchanged work, which only happens in the limit where one of the systems becomes classical \cite{Bertet2002}. This scenario extends and generalizes the recently proposed charging battery protocols~\cite{Alicki2013,Binder2015,Campaioli2017,Andolina2018,Ferraro2018,Andolina2019,Barra2019,Mitchison2020}: it treats work providers and work receivers in a symmetric way within a unified picture, and relates the performance of the charging protocol to nature of energy exchanges between the two systems. It motivates a number of novel investigations, related e.g. to the optimization of energy recovery from one or both systems, the derivation of fundamental bounds relating work and local energy extraction, or the study of how quantum resources impact the bound. Since this framework does not necessarily involves a thermal bath, it was dubbed quantum energetics by some of us \cite{Lea,Alexia-PRX-Quantum2021}.

We then start to explore these questions, conducting our analyzes in the waveguide quantum electro-dynamics (WG-QED) platform, taking a qubit coupled to a waveguide as our bipartite system. WG-QED is an accurate paradigm to investigate light matter interaction in the quantum regime. It provides a complete, analytically solvable description of a driven quantum open system, where the drive, the qubit, and the bath evolve as an isolated system \cite{Weisskopf,Gardiner1985,Fan2010inout,Ciccarello2017,Molmer2019,Fischer} -- which is a major advantage for our present energetic considerations. We show that the work flows, respectively performed by the qubit and the field on one another, compensate each other, singling out for the first time the self-reaction work giving rise to spontaneous emission~\cite{Dalibard,Milonni}. We evidence that the work performed by a coherent pulse on the qubit is always larger than the work that can later be extracted from the qubit, i.e. its ergotropy~\cite{Allahverdyan2004,Ghosh2019,Francica2020}. We dub this bound the classical ergotropy bound, and show that it gets violated with a single-photon pulse -- operating as an energetic witness of non-classicality.  \\

\textit{General framework }--We first introduce our system and notations, and recall the framework introduced in \cite{Weimer2008, Hossein-Nejad2015}. We consider two quantum systems $q$ and $f$ with respective free Hamiltonians $H_{q}$ and $H_{f}$. Throughout the paper, we shall use the interaction picture with respect to $H_q+H_f$, such that the coupling Hamiltonian $V(t)$ is time-dependent. The reduced equation of motion for the system $k\in\left\{ q,f\right\}$ is obtained from the von Neumann-Liouville
equation $d\rho/dt=-\left(i/\hbar\right)\left[V(t),\rho(t)\right]$, with $\rho(t)$ being the joint state of $qf$.
\begin{equation}
\frac{d\rho^{k}}{dt}=-\frac{i}{\hbar}\left[\mathcal{H}^{k}(t),\rho^{k}(t)\right]-\frac{i}{\hbar}\text{Tr}_{l\neq k}\left\{ \left[V(t),\chi(t)\right]\right\}.\label{eq:reduced dynamical equation}
\end{equation}
$\rho^{k}(t)=\text{Tr}_{l\neq k}\left[\rho(t)\right]$
is the reduced state of system $k$ and $\chi(t)=\rho(t)-\rho^{q}(t)\otimes\rho^{f}(t)$
is the correlation matrix at time $t$. We emphasize that this equation holds for any initial state of the systems $q$ and $f$, in particular with initial correlations as well. Eq.~(\ref{eq:reduced dynamical equation}) evidences that the two systems influence
each other in two ways. Namely, each system induces the effective driving
term $\mathcal{H}^{k}(t)=\text{Tr}_{l\neq k}\left[V(t)\rho^{l}(t)\right]$ in the other system reduced dynamics, while a non-unitary term results from the build-up of correlations between the two systems. It is straightforward to show that the former
term is (resp. the latter is not) entropy-preserving. Each term gives rise, in turn, to two distinct energetic changes for the system $k$ that we now characterize. 

From now on, we suppose that $\left[H_{q}+H_{f},V\right]=0$. Throughout the paper, this condition is dubbed ``local energies conservation". This is the case of, e.g. the so-called thermal operations~\cite{Brandao2013}. Moreover, we take the coupling $V(t)$ to be orthogonal to $H_q$ and $H_f$, i.e. $\text{Tr}\left[ V(t)\Pi_i \right]= 0$ for $i\in \{q,f\}$ where $\{\Pi_i\}$ are the projectors on the eigenbasis of $H_i$. In this case the energy of system $k\in\left\{ q,f\right\}$ equals $\mathcal{U}^{k}(t)=\text{Tr}_{k}\left[H_k\rho^{k}(t)\right]$. Substituting Eq.~(\ref{eq:reduced dynamical equation}) into the
energy flow $\dot{\mathcal{U}}^{k}=d\mathcal{U}^{k}/dt$
we obtain 
\begin{eqnarray}
 & \dot{\mathcal{U}}^{k}(t)= & -\frac{i}{\hbar}\text{Tr}_{k}\left\{ \left[H_{k},\mathcal{H}^{k}(t)\right]\rho^{k}(t)\right\} -\frac{i}{\hbar}\text{Tr}\left\{ H_{k}\left[V(t),\chi(t)\right]\right\} \nonumber \\
 & \equiv & \dot{\mathcal{W}}^{k}(t)+\dot{\mathcal{Q}}^{k}(t),
\end{eqnarray}
for $k\in\left\{ q,f\right\} $. We have denoted by $\dot{\mathcal{W}}^{k}$
(resp. $\dot{\mathcal{Q}}^{k}$) the energy flow induced by the
unitary (resp. non-unitary) part of the reduced equation of motion. The former stands for the work flow. It is the only term remaining when one of the systems becomes classical, where it matches the usual definition of work. In the general case where entanglement builds up, this definition treats work providers and receivers in a symmetric way. Unlike in \cite{Weimer2008, Hossein-Nejad2015}, we shall not treat this component ``heat" since it does not correspond to some energy exchange with a thermal bath~\cite{Esposito_2010} nor with a classical measuring device~\cite{Elouard2017}. We thus dub it the correlation energy flow.  When local energies are conserved, we show that $\dot{\mathcal{W}}^{q}+\dot{\mathcal{W}}^{f}=0$ (see~\cite{Suppl}).
This result reveals that the work flows, respectively performed by each system on one another, are equal and can be interpreted as the quantum version of the action-reaction principle. \\

\textit{Operational scenario}--We now propose a scenario providing an operational meaning to the definitions above. The quantum systems $q$ and $f$ are initially in their ground states, see Fig.\ref{fig1}(a)-(c). At $t=t_0$, Alice prepares the pure product state $\ket{\psi^q(t_0)}\otimes\ket{\psi^{f}(t_0)}$, which has a cost $-W_{\text{A}}=\mathcal{U}^{q}(t_0)+\mathcal{U}^{f}(t_0)$.  This energy cost corresponds to work in the sense defined above, since Alice performs unitary operations. In our convention the work is positive when it is extracted from a system.  At time $t=t_0^{+}$, the coupling between the two systems is switched on and they evolve unitarily under the total Hamiltonian up to some time $t$, after which the coupling Hamiltonian 
is switched off. Switching the coupling on and off has no energy cost because of the local energies conservation. This condition implies in particular that the coupling energy remains constant all along the interaction, i.e. $\Delta\mathcal{U}^{q}=-\Delta\mathcal{U}^{f}$. It allows us to analyze the coupled evolution as an exchange of energy between the two systems. 

After switching the coupling off, Bob attempts to extract energy from each system using local unitary operations. As the systems get entangled, their reduced entropies increase, lowering the amount of local work $W_\text{B}$ that Bob can extract, $W_\text{B} \leq -W_\text{A}$. A sufficient condition for reaching the bound at any time is that $q$ and $f$ remain in a product state, i.e. that $q$ or $f$ is initially prepared in a classical state. In quantum optics, the quintessential example of a classical state is a high-intensity coherent field, injected in a bosonic mode~\cite{Cohen1997, Bertet2002}. 

This scenario provides new conceptual tools to analyze battery charging  protocols~\cite{Alicki2013,Binder2015,Campaioli2017,Andolina2018,Ferraro2018,Andolina2019,Barra2019,Mitchison2020}, that usually involve a ``working substance" as a work donor and a ``quantum battery" as a work receiver. The figure of merit to maximize is the final ergotropy of the battery, i.e. the maximal amount of work that can be locally extracted from the battery by local unitary operations~\cite{Allahverdyan2004}. From the analysis above, it appears that a maximal battery ergotropy is reached when the two systems solely exchanged work during the loading step. This invites to relate the performance of the protocol to the amount of work exchanged between the working substance and the battery. Our goal is now to establish tight bounds relating work, correlation energy and local energy extraction, which calls for a more concrete scenario.  \\

\textit{Waveguide QED setup.---} From now on, $q$ and $f$ respectively feature a qubit and a one-dimensional waveguide (1DWG), i.e. a reservoir of electromagnetic modes at zero temperature (see Fig.~\ref{fig1}(d)-(f)). The qubit Hamiltonian reads $H_{q}= \hbar \omega_0 \sigma_{+}\sigma_{-} $, with $\sigma_{-}=\ket{g}\bra{e}$, $\sigma_{+}=\sigma_{-}^{\dagger}$, and $\sigma_z=\ket{e}\bra{e}-\ket{g}\bra{g}$, where $\ket{e}$ (resp. $\ket{g}$) denotes the qubit excited (resp. ground) state. The field Hamiltonian is $H_{f}=\sum_{k=0}^{\infty} \hbar\omega_k b_k^\dagger b_k$, where the operator $b^{\dagger}_{k}$ (resp. $b_k$) creates (resp. annihilates) one photon of frequency $\omega_k \geq 0$ and momentum $\hbar k =\hbar \omega_k v^{-1}$, with $v\geq 0$ being the field's group velocity. The coupling Hamiltonian in the Schr\"odinger picture reads $V = \hbar g\sum_k \left(i b_k \sigma_+ + \text{h.c.}\right)$, where h.c. stands for hermitian conjugate and $g$ is the coupling constant, assumed uniform over the modes. 
The lowering operator at the position $x$ in the interaction picture is given by~\cite{Loudon2000}
\begin{equation} 
\label{eq:a}
a(x,t)=\sqrt{\frac{1}{\cal{D}}}\sum_{k=0}^{\infty} e^{-i \omega_k( t-x/v)}b_k=a(0,t-x/v),
\end{equation}
where $\cal{D}$ is the mode density, taken as uniform. Equation~\eqref{eq:a} evidences that the field operators depend on the single variable $\tau=t-x/v$ and verify the bosonic commutation relations $[a(x,t),a^{\dagger}(x^{\prime},t^{\prime})]=[a(0,\tau),a^{\dagger}(0,\tau^{\prime})]=\delta(\tau-\tau^{\prime})$, with $\tau^{\prime}=t^{\prime}-x^{\prime}/v$. 
The qubit is located at the position $x=0$ of the waveguide, such that the coupling Hamiltonian in the interaction picture reads
 \begin{align} \label{eq:V}
V(t)=i \hbar \sqrt{\gamma} \sigma_{+}(t) a(0,t) + \text{h.c.},
\end{align} 
where $\gamma = g^2 \cal{D}$ is the qubit spontaneous emission rate and $\sigma_{+}\left(t\right)=e^{i\omega_{0}t}\sigma_+$. 

Equations~\eqref{eq:a} and \eqref{eq:V} suggest that the field-qubit unitary dynamics can be modelled as a series of interactions, or collisions, between the qubit and a series of temporal modes that propagate with velocity $v$, see Fig.~\ref{fig1}(d)-(f). Our approach thus bears similarities to the so-called collisional or repeated interactions model~\cite{Ciccarello2017,Landi2019,Strasberg2019}, with the substantial difference that the dynamics is deduced from first principles and provides full access to the joint qubit-field state. In the regions where $x<0$ or $x>0$, a wavepacket travels without deformation from left to right. All field observables can thus be related to the input and output operators $a_{\text{in}}(t)=\lim_{\epsilon\rightarrow0^{-}}a(\epsilon,t)$ and $a_{\text{out}}(t)=\lim_{\epsilon\rightarrow0^{+}}a(\epsilon,t)$, respectively, that satisfy
\beq
a\left(0,t\right)=\frac{1}{2}\left[a_{\text{in}}\left(t\right)+a_{\text{out}}\left(t\right)\right]. \label{eq:separation input-output}
\eeq
Solving the coupled equations of motion for the field and the qubit gives rise to the mean input-output relation (see~\cite{Suppl})
\bea\label{eq:in_out_rel}
\langle a_{\text{out}}(t)\rangle=\langle a_{\text{in}}(t)\rangle-\sqrt{\gamma} \langle \sigma_{-}(t)\rangle.
\eea
Note that Eq.~\eqref{eq:in_out_rel} should not be confused with the textbook input-output relation written in Heisenberg representation, that holds for the operators instead~\cite{Gardiner1985}.

The reduced equation capturing the dynamics of the qubit is presented in~\cite{Suppl}, and involves the effective driving operator $\mathcal{H}^q (t)
=i \hbar\sqrt{\gamma}\langle a(0,t)\rangle  \sigma_{+}(t)+\text{h.c.}$. Interestingly, the injection of a coherent input pulse leads to the same evolution for the qubit observables as the Optical Bloch Equations (OBE) at zero temperature~\cite{Cohen1997,Cyril2020}. The analytical solution provided by the WG-QED model can thus be seen as a purification of the well-known OBE, where the qubit, the drive, and the zero temperature bath evolve unitarily. Given Eqs. \eqref{eq:separation input-output} and \eqref{eq:in_out_rel}, it is straightforward to see that the driving operator $\mathcal{H}^{q}(t)$ involves the field radiated by the qubit $\sqrt{\gamma}\langle \sigma_{-}(t)\rangle$, while it reduces to the input field $\langle a_{\text{in}}(t)\rangle$ in the case of the OBE. This is expected since unlike our framework, the OBE do not provide any quantum description of the electromagnetic field: it is either treated as a classical drive or as a thermal bath. Both descriptions do converge, however, when the field amplitude is large such that $|\langle a_{\text{in}}(t)\rangle|^2 \gg \gamma$~\cite{Suppl}. This defines the classical limit of the field, where stimulated emission in the driving mode largely overcomes spontaneous emission.\\

\textit{Classical ergotropy bound.---} We now introduce the framework needed for our energetic analyzes. Note that the coupling is always on: there is thus no cost related to switching the coupling on and off, simplifying the energetic analyzis. Moreover, we restrict the study to the cases where $\langle V(t_0) \rangle =0$ , which is satisfied if the qubit dipole or the mean input field at $t=t_0$ are zero, or if both have the same phase. We show in~\cite{Suppl} that if the field is resonant with the qubit, then the mean coupling remains zero all along the evolution -- which realizes the local energies conservation condition. This allows us to define a unique work flow $\dot{\mathcal{W}}=\dot{\mathcal{W}}^{q}=-\dot{\mathcal{W}}^{f}$, that features the work received by $q$ and provided by $f$.

We employ Eqs.~\eqref{eq:separation input-output}  and \eqref{eq:in_out_rel}  to expand the work flow $\dot{\mathcal{W}}$ as a function of the input field and qubit dipole (see~\cite{Suppl}). We obtain $\dot{\mathcal{W}}(t)=\hbar \omega_0 \left(2\sqrt{\gamma}\text{Re}[\langle \sigma_{-}(t)\rangle \langle a_{\text{in}}(t)\rangle^*]- \gamma|\langle \sigma_{-}(t)\rangle|^2\right)$. The first term features work exchanges by stimulated emission~\cite{Cottet2017,Valente2017,Landi2019,Cyril2020} and is the only term remaining in the classical limit of the field. The latter is remarkable since it remains in the absence of input field, $\left\langle a_{\text{in}}\left(t\right)\right\rangle =0$, i.e. during spontaneous emission. From the qubit perspective, it features the ``self-work" performed by the radiation reaction force~\cite{Dalibard,Milonni}. This self-work is always negative since the qubit can only provide energy to the vacuum field. From the field perspective, it corresponds to the ``spontaneous work" already evidenced in~\cite{Juliette2020}. It is interesting to rewrite the work flow as a function of the field quantities only: $\dot{\mathcal{W}}(t)=- \hbar \omega_0 \left(|\langle a_{\text{out}}(t)\rangle|^2-|\langle a_{\text{in}}(t)\rangle|^2\right)$. This equation reveals that the work is the energy change of the field coherent component \cite{Scully1999}, which is accessible through homodyne or heterodyne measurement schemes~\cite{Wiseman2009,Cottet2017}. It evidences that the amount of exchanged work is fully encoded in the field state, from which it can be directly measured -- in other words, work is an observable. Conversely, the energy change of the field incoherent component directly reflects that the qubit and the field got correlated during their interaction.

Our framework gives rise to a general energetic bound if the pulse statistics is coherent. Our bound involves the so-called qubit ergotropy, that features the maximal amount of work that can be extracted from the qubit by unitary operations~\cite{Allahverdyan2004}. Its expression is computed in \cite{Suppl} and reads $\mathcal{E}^{q}(t)= \hbar\omega_0\left[\langle\sigma_z(t)\rangle+r(t)\right]/2$, with $r(t)=\sqrt{\langle\sigma_x(t)\rangle^2+\langle\sigma_y(t)\rangle^2+\langle\sigma_z(t)\rangle^2}$. Then, for any initial state of the qubit, the correlation energy flow reads $\dot{\mathcal{Q}}(t)=\hbar \gamma(|\langle \sigma_{-}\rangle|^2-\langle \sigma_{+}\sigma_{-}\rangle)\leq0$~\cite{Suppl}, such that the correlation energy is always negative. In turn, this implies that $\mathcal{W}(t)\geq \Delta \mathcal{E}^{q}(t)$, where $\Delta \mathcal{E}^{q}(t)= \mathcal{E}^{q}(t)-\mathcal{E}^{q}(t_0)$ is the change in the qubit ergotropy~\cite{Suppl}. This inequality reveals that a coherent field cannot provide more ergotropy than work to the qubit. Henceforth, we refer to this bound as the classical ergotropy bound. It is reached at any time if no correlations are created, i.e. if the field state reaches the classical limit.

The classical ergotropy bound is violated if the input field is a resonant single-photon pulse, a paradigmatic example of non-classical statistics. A particularly striking case is provided by a mode-matched inverted exponential. Such fields have been theoretically~\cite{Fan2010,Wang2011} and experimentally~\cite{Golla2012,Bader2013,Aljunid2013} shown to lead to complete population inversion (see~\cite{Suppl}), yielding ergotropy to the qubit as soon as $\langle \sigma_z \rangle \geq 0$. Meanwhile, a single-photon pulse has no coherent component, such that $\langle a(0,t) \rangle=0$ at any time. Therefore, it does not perform any work on the qubit, leading to ${\cal W}(t)=0 \leq \Delta {\cal E}^q(t)$ when the population gets inverted. This violation of the classical ergotropy bound established above provides an energetic witness to probe the non-classicality of the field. This remarkable feature completes other criteria based on Wigner function negativities~\cite{Scully1999,Bertet2002} and relies on operational quantities that are experimentally accessible.\\

\textit{Impact of quantum resources on work recovery.---} We now propose an experimentally realistic scenario to probe the bound derived above. To do so, it is convenient to make the connection with the general framework presented in the first part of the paper. The field and the qubit are initially in their ground state. At $t_0 \leq0$, Alice prepares an input light pulse while the qubit remains in $\ket{g}$. Under the free evolution, the light pulse propagates from the negative to the positive $x$ axis, see Fig.~\ref{fig1}(d)-(f). It typically interacts with the qubit when it crosses $x=0$, at a time chosen as the origin $t=0$. At long times with respect to the pulse duration and the spontaneous emission time, the qubit has relaxed back to its ground state and the field is in a pure state defined by the scattered pulse. 

\begin{figure}[t]
	\begin{center}
\includegraphics[width=0.95\columnwidth]{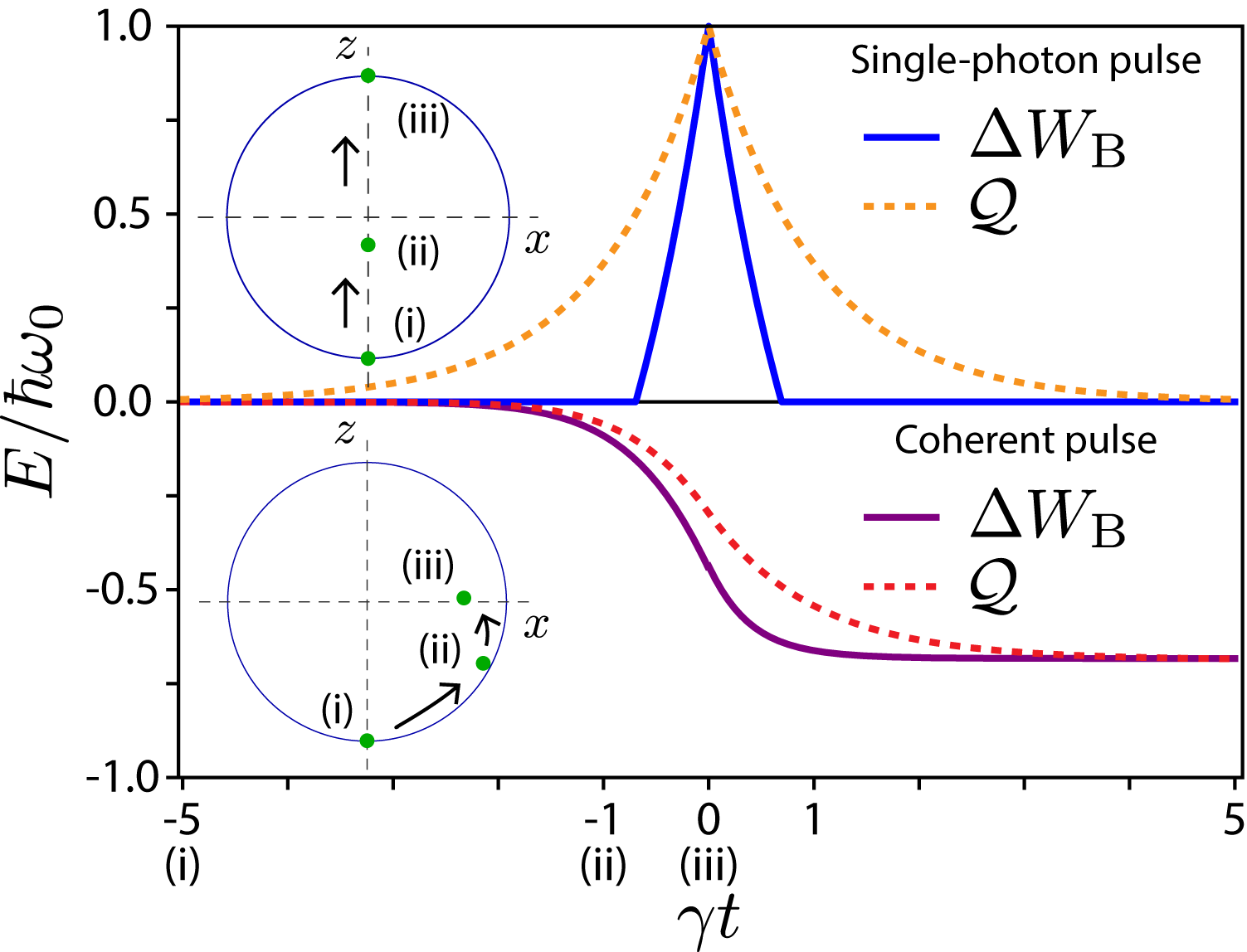}
\caption{Classical ergotropy bound and correlation energy. At $t=t_0$ the qubit is prepared in $\ket{g}$ and a light pulse at the position $x=-vt_0$. The pulse envelope reads $\alpha\left(t\right)=\sqrt{\gamma}e^{\gamma t/2-i\omega_{0}t}$ for $t\in\left[-\infty,0\right]$ and $\alpha\left(t\right)=0$ for $t\in(0,\infty]$ (see~\cite{Suppl}). The extra-amount of work $\Delta W_{\text{B}}(t)$  (See text) is plotted for the coherent field case (solid purple) and the single-photon case (solid blue). The correlation energy $\mathcal{Q}(t)$ is plotted for the coherent (dashed red) and single-photon (dashed orange) cases. We set $t_0=-5/\gamma$. 
\label{classical_bound}
}
\end{center}
\end{figure}

At any time $t$, we evaluate the capacity of Bob to locally extract the respective energies of the field and the qubit. Here we assume that Bob can only employ a classical resonant field to generate his unitary operations. He can thus fully extract the qubit ergotropy by optimizing the phase and duration of the driving pulse. Conversely, he can only displace the light field, hence extract the energy of the field coherent component. This can be done, e.g. by using homodyning schemes~\cite{Wiseman2009,Cottet2017}. The complete ergotropy of the light pulse is thus not recovered by this operation. As we show below, this experimental constraint on the class of operation Bob can perform provides a direct access to the classical ergotropy bound.

The maximum amount of work that Bob can extract at time $t$ equals $W_{\text{B}}(t)={\cal E}^{q}(t)+{\cal E}^{f}_{coh}(t)$. We have introduced ${\cal E}^{f}_{coh}(t) = \hbar \omega_0\left( \int_{t_0}^{t}dt'|\langle a_{\text{out}}(t')\rangle|^2+\int_{t}^{\infty}dt'|\langle a_{\text{in}}(t')\rangle|^2\right)$ the total energy of the field coherent component \cite{Scully1999}, whose change in time equals the work performed by the qubit. We now consider the extra amount of work $\Delta W_{\text{B}}(t)=W_{\text{B}}\left(t\right)-W_{\text{B}}\left(t_0\right)$ that Bob can extract at time $t$ with respect to $t=t_0$. It can also be written as $\Delta W_{\text{B}}(t)=\Delta \mathcal{E}^{q}\left(t\right)-\mathcal{W}\left(t\right)$. $\Delta W_{\text{B}}(t)$ thus features a measurable quantity that captures our bound. It is plotted on Fig.~\ref{classical_bound} in the two situations studied above: a coherent pulse containing one photon on average and a single-photon wavepacket. In the first case, the classical ergotropy bound translates into the decrease of $\Delta W_{\text{B}}$. Conversely, for the single-photon pulse, the bound violation induces an enhancement of the energy locally extracted by Bob. We have also plotted the correlation energy ${\cal Q}$ in both cases, which gives insights into the physical origin of the bound. As it appears on the figure, a coherent field only gives rise to a negative correlation energy, while correlations induced by the quantum pulse can efficiently fuel the qubit. \\

\textit{Outlooks.---} Our work brings new evidences that quantum resources impact the energetic behavior of quantum systems, providing a new ``quantum energetic signature"~\cite{Uzdin2015}. These new mechanisms can take the form of operational energetic witnesses, that complement the well-known quasi-probability distributions of quantum optics. The experiments we propose are feasible on state-of-the-art platforms of integrated photonics~\cite{Loredo2019}, superconducting circuits~\cite{Gu2017} and atomic physics~\cite{PRLRempe}. Beyond the WG-QED scenario, this framework opens new avenues to extend the concept of work in the quantum world -- that can be further used to analyze other kinds of interactions, e.g. modeling measurement processes. The energy recovery scenario could find useful applications in the realm of quantum technologies, e.g. the work spent to run quantum circuits could be optimally recovered when the algorithm is completed.\\

\textit{Acknowledgments.---} The authors warmly thank G. Landi for his precious comments. This work was supported by the Foundational Questions Institute Fund (Grant number FQXi-IAF19-05), the Templeton World Charity Foundation, Inc (Grant No. TWCF0338) and the ANR Research Collaborative Project ``Qu-DICE" (ANR-PRC-CES47).

\bibliography{Thermo_wg_bibliography.bib}

\begin{thebibliography}{53}
\bibitem [{Sup()}]{Suppl}%
  \BibitemOpen
  \href@noop {} {\bibinfo {title} {See {S}upplemental {M}aterial at [{URL} will
  be inserted by publisher] for further details.}}\BibitemShut {Stop}%
\end{thebibliography}

\end{document}